\documentclass[%
 reprint,
 amsmath,amssymb,
 aps,
]{revtex4-1}

\usepackage{array,longtable}
\newcolumntype{L}{>{\tiny $}p{0.33\columnwidth}<{$}}
\newcolumntype{M}{>{\scriptsize $}p{0.33\columnwidth}<{$}}
\newcolumntype{N}{>{\scriptsize $}p{0.43\columnwidth}<{$}}
\usepackage{ulem}
\usepackage{graphicx}
\usepackage{dcolumn}
\usepackage{bm}
\usepackage{subfigure}

\newcommand{\bs}[1]{\boldsymbol{#1}}

\usepackage{physics}
\usepackage{float}
\usepackage{hyperref}

\usepackage{color}

\usepackage{braket}

\begin{document}


\title{Competing orders in the Hofstadter $t$--$J$ model}

\author{Wei-Lin Tu$^{1,2,3}$, Frank Schindler$^{4}$, Titus Neupert$^{4}$, and Didier Poilblanc$^{2}$}

\affiliation{
$^{1}$\textit{Department of Physics, National Taiwan University, Daan Taipei 10617, Taiwan}\\ 
$^{2}$\textit{Laboratoire de Physique Th$\acute{e}$orique, IRSAMC, Universit$\acute{e}$ de Toulouse, CNRS, UPS, France}\\
$^{3}$\textit{Institute of Physics, Academia Sinica, Nankang Taipei 11529, Taiwan}\\
$^{4}$\textit{Department of Physics, University of Zurich, Winterthurerstrasse 190, 8057, Zurich, Switzerland}
}

\date{\today}

\begin{abstract}
The Hofstadter model describes non-interacting fermions on a lattice in the presence of an external magnetic field. Motivated by the plethora of solid-state phases emerging from electron interactions, we consider an interacting version of the Hofstadter model including a Hubbard repulsion $U$. We investigate this model in the large-$U$ limit corresponding to a $t$--$J$ Hamiltonian with an external 
(orbital) magnetic field. By using renormalized mean field theory supplemented by exact diagonalization calculations of small clusters, we find evidence for competing symmetry-breaking phases, exhibiting (possibly co-existing) charge, bond and superconducting orders. Topological properties of the states are also investigated and some of our results are compared to related experiments involving ultra-cold atoms loaded on optical lattices in the presence of a synthetic gauge field. 

\end{abstract}

\pacs{Valid PACS appear here}
\maketitle


\section{\label{sec:level1}Introduction}\label{sec:introduction}

The Hofstadter butterfly alongside with its Hamiltonian, the Harper-Hofstadter Hamiltonian \cite{Hofstadter}, serves as basis for the study of non-interacting lattice fermions moving in an orbital magnetic field. With the increasing accuracy of experiments, e.g., in laser-manipulated cold atom systems in a two-dimensional square lattice \cite{Aidelsburger,Miyake,Kennedy,Aidelsburger2,Mancini,Stuhl}, it becomes possible to investigate minute details of this non-interacting model. In addition, cold atom systems have proven to be able to emulate interacting fermionic or bosonic systems \cite{Kennedy,Cooper,Goldman,Lacki}, which may lead to the realization of exotic material phases such as a cold-atom analogue of the fractional quantum Hall (FQH) effect \cite{Hansson}, as suggested by promising results from exact diagonalization (ED) of small clusters~\cite{Sorensen,Hafezi,Moller,Sterdyniak}.

Another motivation to study the square lattice in the presence of orbital magnetic fields and strong correlations comes from 
the field of high-$T_{\mathrm{c}}$ superconductivity.
The Hubbard Hamiltonian on the square lattice (without external flux) was meant to explain the mechanism of high-$T_{\mathrm{c}}$ superconductivity by introducing an on-site interaction $U$, which leads to Mott physics \cite{Anderson}. A $t$--$J$ Hamiltonian arises from the Hubbard model when the interaction becomes large compared to the bandwidth, with $J=4t^2/U$ being the antiferromagnetic (AF) coupling between nearest-neighbor spins (and $t$ being the hopping). 
In Anderson's original resonating valence bond (RVB) scenario, superconductivity emerges by doping the parent Mott insulator away from half-filling, and proposals for different Mott spin liquid phases have been given. One of them is the Affleck-Marston half-flux state \cite{Affleck,Marston,Kotliar}, which can be mapped onto free electrons on a lattice with half a magnetic flux quantum per plaquette (and effective hopping $J$). Away from half-filling, the (mean-field) Affleck-Marston flux phase acquires lowest energy density when the flux per unit cell equals exactly the fraction $\nu=\frac{1}{2}(1-\delta)$ where $\delta$ is the doping level~\cite{Lederer,Nori}. In fact, the corresponding interacting states can be viewed as a Gutzwiller projection of the free fermionic wavefunctions under magnetic flux.  This reveals important aspects of the RVB physics and thus motivates us to perform calculations directly with the $t$--$J$ Hamiltonian in the presence of an {\it actual} external magnetic flux, as we do in the present study. 

In fact, numerous different phases have been obtained by investigating Hubbard and $t$--$J$ type Hamiltonians in attempts to find the proper theoretical description of high-$T_{\mathrm{c}}$ cuprates~\cite{Tu, Choubey, Zhang, Himeda, Ogata, Christensen, Chou, Chou2, Poilblanc2, White, White2}, revealing 
low-energy intertwined inhomogeneous states, such as stripes, bidirectional charge-ordered states, checkerboard
patterns, and so on. Recently, tensor network studies~\cite{Corboz} and density matrix embedding theory~\cite{BZheng} provided new evidence that the ground state (GS) of the Hubbard model could indeed be inhomogeneous at finite doping and that its phase diagram shows co-existence of $d$-wave superconducting (SC) order with other 
instabilities. This fact hinders the possible  emergence of topologically non-trivial phases since the latter compete 
with instabilities. 
However, in the presence of an external orbital magnetic field, flat bands formed as Landau levels re-introduce this possibility. 
Also from this perspective it is therefore interesting to consider orbital effects by studying the $t$--$J$ Hamiltonian in presence of an orbital magnetic field.

\begin{table*}[t]
\begin{center}
\resizebox{1.1\columnwidth}{!}{
\begin{tabular} {| c | c | c | c | c | c | c | c | c | c |}
\hline
$\rho$ & $\Phi$   &$\nu/\nu^*$ &  $N_{\mathrm{s}}$ & $N_{\mathrm{e}}$ & $N_\Phi $  & $S$ & Unit Cell & Instabilities\\ \hline
\hline
$7/16$ & $7/16$  &$1$  &  16$\times$16 & 224 &112  & $0$ & 1$\times$1 & None \\ \hline
 $7/16$ & $5/16$ &$7/5$   & 16$\times$16 & 224 &80  & $0$ & 2$\times$2 & BDW/PDW \\ \hline
 $7/16$ & $3/16$ & $7/3$  & 16$\times$16 & 224 &48  & $0$ & 4$\times$4 & CDW, BDW/PDW\\ \hline
$7/16$ & $1/16$   &$7$  &  16$\times$16 & 224 &16  & $0$ & $\sqrt{2}$$\times$$\sqrt{2}$ & SC \\ \hline
$7/32$ & $7/16$   & $1^*$  & 12$\times$12 & 63 &63  & $FP$ & 1$\times$1 & None \\ \hline
 $1/8$ & $1/4$   & $1^*$  &12$\times$12 & 36 &36   & $FP$ & 2$\times$2 & CDW, BDW\\ \hline
$1/8$ & $7/16$   & $4/7^*$  & 12$\times$12 & 36 &63   & $FP$ & 1$\times$1 & None \\ \hline
$1/16$ & $5/16$   &$2/5^*$  &  12$\times$12 & 18 &63   & $FP$ & 4$\times$4 & CDW, BDW \\ \hline
$1/16$ & $7/16$   &$2/7^*$  &  12$\times$12 & 18 &45   & $FP$ & 1$\times$1 & None\\ \hline
\end{tabular} }
\caption{Parameter sets used in Sections~\ref{sec:CFP}, ~\ref{sec:FP}, and \ref{sec:FQH}.
$N_{\mathrm{s}}$, $N_{\mathrm{e}}$, and $N_\Phi$ are the site, electron and flux numbers used for performing RMFT 
(those for the ED on a $4\times 4$ cluster are obtained from a simple rescaling). Sets are listed with decreasing electron filling 
from top to bottom. The GS is either a singlet ($S=0$) or fully polarized (FP), i.e., the total spin is $S=\frac{N_{\mathrm{e}}}{2}$ (in that case $\nu^*=2\nu$ is listed and marked with an asterisk). The supercell associated to a
possible spontaneous (charge or bond) ordering is also shown. 1$\times$1 means the GS is uniform. CDW, BDW, and PDW stand for charge, bond, and pairing density wave. SC means staggered current modulation. For $\rho=7/16$ and $\Phi=5/16$ or $3/16$, including (d-wave) superconducting order in addition to
CDW/BDW order gives a PDW self-consistent solution with lower energy. For $\rho=1/8$ and $\Phi=1/4$ ($\nu^*=1$), the $2\times 2$ 
modulation is induced by a staggered
potential (Section \ref{sec:FQH}). Otherwise, translation symmetry breaking (if any) occurs spontaneously.}
\label{tab:parameters}
\end{center}
\end{table*}

By no means is there a single analytic or numerical methodology for solving the $t$--$J$ Hamiltonian. Here, we will apply two complementary approaches. One is the renormalized mean-field theory (RMFT) first proposed by Zhang and Rice \cite{Zhang} with further revision in Refs.~\cite{Himeda,Ogata,Yang3,Christensen} to include second order (bond) renormalization when spin polarization is present. This method, as any mean field technique, can only detect symmetry-broken phases provided the proper order parameters are introduced by hand, but allows to reach large system sizes. We compare our results to ED calculations, which are a priori unbiased, but strongly limited in terms of available system sizes. Recently, Gerster \textsl{et al.} \cite{Gerster} demonstrated the existence of a FQH phase akin to the $\nu=1/2$ Laughlin state for the spinless bosonic Harper-Hofstadter model by using a tree-tensor network ansatz. This shows that it is possible to obtain novel quantum phases from the Hofstadter Hamiltonian in the presence of interactions and, therefore, provides 
another motivation to study this model with {\it spinful} fermions.

This work is structured as follows. In Sec.~\ref{sec:setting}, we present our model Hamiltonian and the parameter sets for the phases we found. Also, the numerical methods we applied will be briefly explained while the details are included in the Appendix. In Sec.~\ref{sec:CFP}, we will revisit the commensurate flux phase (CFP), which has been studied in previous work~\cite{Poilblanc, Poilblanc2}. Here, we will in particular focus on charge instabilities and topological features of the CFP. Instabilities towards ferromagnetic phases (fully polarized states) are described in Sec.~\ref{sec:fully polarized}, showing good agreement between our two numerical approaches. Topological aspects (e.g., the computation of Chern numbers) and comments on the search for potential FQH physics are subsumed in Sec.~\ref{sec:FQH}, followed by the conclusion in Sec.~\ref{sec:conclusion}.

\section{\label{sec:level1}Model Hamiltonian, methods and parameter sets}\label{sec:setting}

Here, we consider the 2D $t$--$J$ model, i.e., the large-$U$ limit of the 2D Hubbard model, in an external magnetic field as our interacting Hamiltonian,
\begin{equation}
\begin{aligned}
\label{eq:H}
H&=\underbrace{-\sum_{\langle i,j \rangle, \mu} P_{\mathrm{G}} \left(t_{ij} c^\dagger_{i \mu} c_{j \mu} +\mathrm{h.c.} \right) P_{\mathrm{G}}}_{H_\text{kin}} \underbrace{+\sum_{\langle i,j \rangle}J \boldsymbol{S}_i \cdot \boldsymbol{S}_j}_{H_\text{pot}},
\\ & \quad t_{ij}=t \, e^{\mathrm{i} A_{ij}}=t_{ji}^*, \quad \boldsymbol{S}_i = \sum_{\mu,\nu} c^\dagger_{i\mu} \boldsymbol{\sigma}_{\mu \nu} c_{j\nu},
\end{aligned}
\end{equation}
where $c^\dagger_{i \mu}$ ($c_{i \mu}$) is the creation (annihilation) operator for an electron of spin $\mu = \uparrow, \downarrow$ on lattice site $i$, so that $n_{i \mu} = c^\dagger_{i \mu} c_{i \mu}$ is the site number operator per spin, $P_{\mathrm{G}} = \prod_i (1 - n_{i \uparrow} n_{i \downarrow})$ is the Gutzwiller projector onto the Hilbert subspace of at most singly-occupied sites, and 
$\boldsymbol{\sigma} = (\sigma_x, \sigma_y, \sigma_z)^\mathsf{T}$ is the vector of $2 \times 2$ Pauli spin matrices. In the exact mapping from Hubbard to $t-J$ model there is another term of order $t^2/U$, the so called three-site hopping, which describes hopping of singlet pairs. This term has been shown to have no influence on the mean-field phase diagram \cite{Ercolessi} and is therefore excluded in our work. The AF coupling $J$ is chosen to be equal to $0.3$ times the hopping $t$ throughout the paper.

The magnetic field enters via the phases $A_{ij} = \int_{i}^{j} \boldsymbol{A} (\boldsymbol{x}) \cdot d\boldsymbol{x}$, where the vector potential $\boldsymbol{A}(\boldsymbol{x})$ is defined by the relation $\boldsymbol{B} (\boldsymbol{x})= \nabla \wedge \boldsymbol{A}(\bs{x})$, corresponding to a flux per plaquette $F = \int_{\framebox(3,3){}} \bs{B} (\bs{x})\cdot d\bs{\Sigma} = A_{i,i+\hat{x}} + A_{i+\hat{x},i+\hat{x} + \hat{y}} + A_{i+\hat{x} + \hat{y},i+\hat{y}} + A_{i+\hat{y},i}$, which we take to be independent of $i$. Here we choose $F = 2\pi \Phi$, with $\Phi$ given by fractions such as $\frac{7}{16}$, $\frac{5}{16}$, etc. Note that since we work in units where $\hbar = e = 1$, $\Phi = 1$ corresponds to one magnetic flux quantum.

The standard procedure of RMFT is to first replace the Gutzwiller projection operator by renormalized factors $g^t$ and $g^s$ so that 
\begin{equation}
\begin{aligned}
\label{eq:H}
&\langle \Psi | c^\dagger_{i\mu}c_{j\mu} | \Psi \rangle=g^t_{ij\mu}\langle \Psi_0 | c^\dagger_{i\mu}c_{j\mu} | \Psi_0 \rangle,\\
&\langle \Psi | \bs{S}_i\cdot \bs{S}_j | \Psi \rangle=g^s_{ij}\langle \Psi_0 | \bs{S}_i\cdot \bs{S}_j | \Psi_0 \rangle,
\end{aligned}
\end{equation}
where $| \Psi_0 \rangle$ is the un-projected wavefunction and $| \Psi \rangle=P_{\mathrm{G}}| \Psi_0 \rangle$. We then transform the four-body operator $\bs{S}_i \cdot \bs{S}_j$ into a quadratic term in $c$ and $c^\dagger$ just as in the standard mean-field process. The final mean-field Hamiltonian is solved iteratively until the desired convergence is achieved. Details are included in the Appendix. This method has been proven useful in the search for high-$T_{\mathrm{c}}$ superconductivity in the $t$--$J$ model (with no flux)~\cite{Tu,Choubey} and, therefore, we shall also adopt 
it to investigate the $t$--$J$ Hamiltonian in the presence of an applied flux.
Our RMFT results will always be compared with Lanczos ED of small clusters, which has been carried out for exactly the same values of
flux and electron density.

Before going further it is useful to add some comments and words of caution regarding the interpretation of the RMFT results.
RMFT is essentially designed to provide an efficient construction of an optimized correlated ansatz to approximate the 
targeted GS of a many-body Hamiltonian, by Gutzwiller projecting a self-consistent non-interacting wavefunction.
It is also accurate in computing GS expectation values, like the energy, and (spontaneous) symmetry breaking at the level of the
RMFT Hamiltonian will translate immediately into similar symmetry breaking of the correlated Gutwiller projected wavefunction.
However, one should refrain from giving a too strong physical meaning to the RMFT spectrum (the spectrum of a simple quadratic Hamiltonian) 
which is not 
guaranteed to be in one-to-one correspondence with the actual many-body excitation spectrum. In particular, it is likely that such a correspondence breaks down completely when approaching the half-filled Mott insulating phase when interactions become essential. 
In short, the RMFT is good to construct the GS manifold but not beyond.

Table~\ref{tab:parameters} shows the parameter sets we have used in the RMFT self-consistent calculations. For simplicity, we choose to work on a square lattice geometry with periodic boundary conditions and a $4\times 4$ {\it magnetic} sublattice is used to encode an integer number of flux quanta. Hence, the flux per plaquette can be chosen as $\Phi = p/q$ with $q=16$ and $p$ any integer, giving a total number $N_\Phi=\Phi N_{\mathrm{s}}$ of magnetic flux quanta piercing the whole torus surface, where $N_{\mathrm{s}}$ is the number of lattice sites. The particle filling $\rho$ is equal to $\frac{N_{\mathrm{e}}}{2N_{\mathrm{s}}}$, with $N_{\mathrm{e}}$ being the number of electrons. The doping with respect to the half-filled Mott insulator
is $\delta=2(\frac{1}{2}-\rho)$. Because of particle-hole symmetry we can restrict to $\delta>0$.
The filling fraction $\nu=\rho/\Phi$ indicates the ratio of Landau levels filled in the corresponding 
non-interacting picture. Clearly it is relevant for zero-magnetization systems, denoted by $S=0$ in Table~\ref{tab:parameters}. In contrast, a fully polarized (FP) GS would instead be ``adiabatically'' connected
to a non-interacting (spinless) fermion system at filling fraction $\nu^*=2\nu=2\rho/\Phi$. The last column of Table~\ref{tab:parameters} contains the information about the unit cell characterizing a possible (spontaneous) ordering for each state. 
Notice that the largest cluster size that can be reached with ED is $4\times 4$ corresponding to a unique magnetic unit cell. In that case, the corresponding flux and electron numbers $N_\Phi=16\times \Phi$
and $N_{\mathrm{e}}=32\times\rho$ need to be integers. In the two following sections, we shall review the properties of the various phases found,
the uniform and modulated flux states (Sec.~\ref{sec:CFP}) and the ferromagnetic FP phases (Sec.~\ref{sec:FP}),
as can be inferred from the properties listed in the last two columns of Table~\ref{tab:parameters}. 

\section{\label{sec:CFP}Uniform and modulated singlet flux phases}\label{sec:CFP}

The first phase of interest which could be realized in this Hamiltonian is the Anderson, Shastry, and Hristopoulos (ASH) state~\cite{Anderson2}. It is also called CFP because of its commensurability condition between the flux and electron filling~\cite{Poilblanc}. It has been shown that these states can be formally written in the quantum spin liquid form, the singlet bond amplitudes of which break the lattice translational symmetry~\cite{Poilblanc3}, and their order of commensurability with the lattice unit length is closely related to the hole density \cite{Lederer,Nori,Poilblanc3}. The stability of the CFP with varying flux, first discussed in Refs.~\cite{Lederer,Poilblanc}, will be revisited here.

\begin{figure}[t]
\centering
\includegraphics[width=0.4 \textwidth]{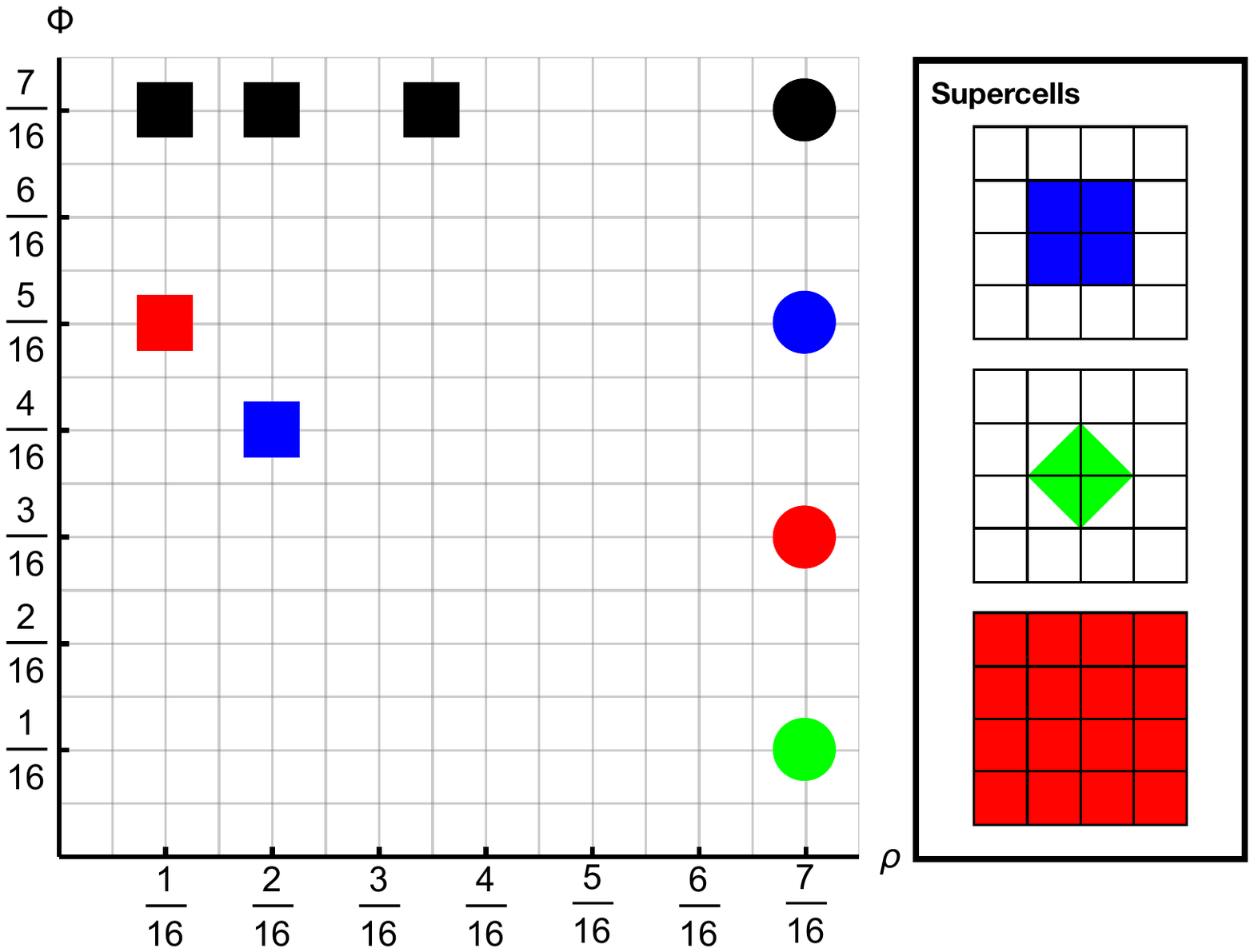}
\caption{``Phase diagram'' vs electron filling $\rho$ and magnetic flux $\Phi$ showing the various phases presented in Table~\ref{tab:parameters}. Circles are non-polarized (singlet) states while squares represent ferromagnets. Black symbols correspond to uniform solutions. Red, green, and blue symbols
encode symmetry-breaking supercells of size $4\times4$, $\sqrt{2}\times \sqrt{2}$, and $2 \times 2$ (with staggered potential for $\Phi=1/4$) respectively.
}
\label{fig: phases}
\end{figure}

\begin{figure}[t]
\centering
\includegraphics[width=0.45 \textwidth]{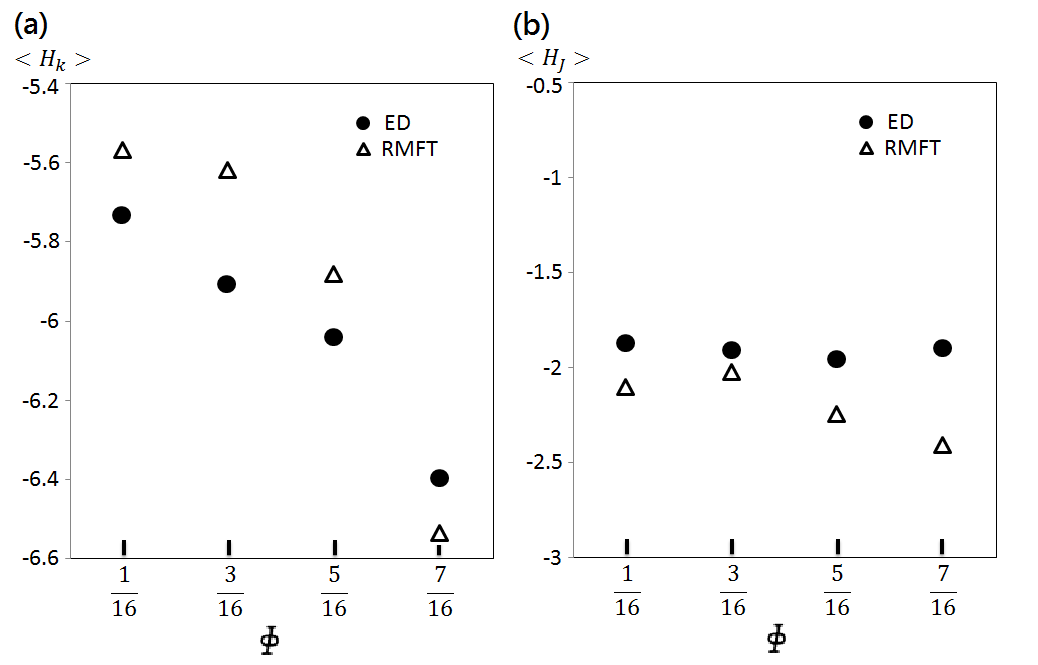}
\caption{Comparison between RMFT and ED energies (per magnetic $4\times 4$ unit cell). (a) Kinetic energy and (b) magnetic (potential) energy vs inserted flux $\Phi$. 
The doping level is fixed to $\delta=1/8$ and $J=0.3t$. The numerical values are given in the Appendix.}
\label{fig: energycomp}
\end{figure}

In this section we fix the electronic fraction to be $\rho=7/16=0.4375$
and study how the states evolve with changing flux. 
This corresponds to a weakly-doped Mott insulator with a doping $\delta=2(\frac{1}{2}-\rho)=1/8$, i.e., two holes per  
magnetic $4\times 4$ supercell.
Within this choice of parameters, a {\it uniform} CFP has only been found for $\Phi=\rho=7/16$ (first line of Table~\ref{tab:parameters}).
For the same doping and other commensurate values of the flux, $\Phi=p/16\ne \rho$ with $p$ an odd integer, singlet phases 
exhibiting lattice symmetry breaking patterns have been found, as is the case for the parameters corresponding to the second,
third and fourth lines of Table~\ref{tab:parameters}. These patterns could correspond to a modulation of the (site) charge density and/or a modulation of the (real) bond hopping amplitude, called here charge density wave (CDW) or bond density wave (BDW),
respectively. CDW and BDW orders may or may not coexist (compare second and
third lines of Table~\ref{tab:parameters}). Staggered current (SC) patterns can also appear without CDW/BDW orders
as described later on (see fourth line of Table~\ref{tab:parameters}).
 
Let us first examine the case $\Phi=\rho$.  The results obtained for $J=0.3t$ ($t=1$) show a homogeneous state and the RMFT band structure 
reveals a large band gap at the chemical potential. This corresponds to a mean-field (unprojected) state where the first Landau level is exactly 
filled.
In the time-reversal symmetry broken state we may calculate the current for each bond as $J_{ij}=g^t_{ij\uparrow}\mathrm{Im}(\chi_{ij\uparrow}e^{i\phi_{ij}})+g^t_{ij\downarrow}\mathrm{Im}(\chi_{ij\downarrow}e^{i\phi_{ij}})$ while the charge hopping is $g^t_{ij\uparrow}\mathrm{Re}(\chi_{ij\uparrow}e^{i\phi_{ij}})+g^t_{ij\downarrow}\mathrm{Re}(\chi_{ij\downarrow}e^{i\phi_{ij}})$, where $\chi_{ij\sigma}=\langle c^\dagger_{i\sigma}c_{j\sigma}\rangle$. (The values of $\phi_{ij}$ at each bond for different $\Phi$ are shown in the Appendix.) For $\Phi=\rho$ all the bonds have zero current, confirming the homogeneous character of this state within the mean-field approach. The energy difference between RMFT and ED is mainly due to the magnetic energy, that of RMFT being smaller than the ED, which also agrees with previous results~\cite{Poilblanc}. 

It has been shown previously that, at fixed doping level $\delta=1/8$, the CFP exhibits an absolute minimum of the magnetic energy at $\Phi=7/16$
corresponding to the exact condition $\Phi=\rho$. However, after adding the competing kinetic energy, the total energy was found to
be lower for a smaller commensurate flux, at least at intermediate values of $J/t$~\cite{Poilblanc}. However, in Ref.~\cite{Poilblanc} 
a simple $t$-$J$ Hamiltonian with {\it no} applied flux was considered, the flux entering only at the level of the projected CFP ansatz. Also, Ref.~\cite{Poilblanc} did not take into account the possibility of CDW/BDW instabilities as well as the more sophisticated form of the Gutzwiller renormalization factors, both of which we have included here. When changing the inserted flux to $\Phi=\frac{5}{16}, \frac{3}{16}$, and $\frac{1}{16}$, the difference of the RMFT and ED magnetic energies becomes smaller as can be seen in Fig.~\ref{fig: energycomp}. 
In contrast to Ref.~\onlinecite{Poilblanc}, where the minimum of the kinetic energy was found at $\phi=\frac{1}{16}$, we find here 
with RMFT that it occurs  at $\phi=\frac{7}{16}$, as for the magnetic part. 
This leads to a robust minimum of the total energy vs.\ flux profile and also generalizes to the case of
 the Affleck-Marston phase for which the minimal energy is found at $\Phi=\rho=1/2$.

Notably, for $\Phi=\frac{1}{16}$ and the same doping $\delta=1/8$, $\nu$ is equal to $7$ which is also an integer, signifying that the first $7$ Landau levels (of the mean-field spectrum) are filled. For this case, the real space pattern revealed by RMFT shows a staggered flux state with homogeneous current on each bond, that is, the current circulation directions are opposite between neighboring plaquettes. The reason is that again an integer number of Landau levels has been filled and the large band gap excludes the possibility of inhomogeneous modulation. Hence, it becomes clear that, for integral $\nu$, the band gap is large enough to suppress the lattice instability.  The integer $\nu$ states are then adiabatically connected to band insulators, and we believe this scenario is generic beyond the 
two cases we have tested here. 

\begin{figure}[t]
\centering
\includegraphics[width=0.45 \textwidth]{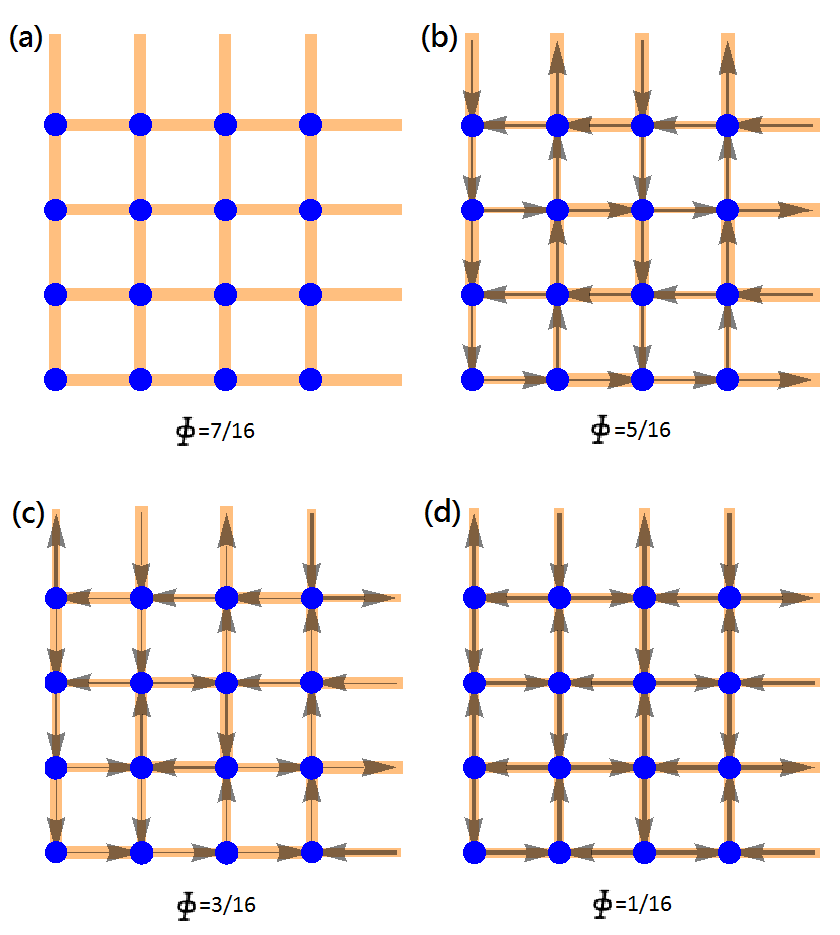}
\caption{Schematic patterns and results for the states in Sec.~\ref{sec:CFP}. (a)-(d) show the current and hopping patterns of each state within the $4\times4$ sublattice. The widths of the underlying orange bars and black arrows represent the magnitudes of hopping and current on each bond separately. The flows of current are indicated by the arrow directions. The numerical values are shown in Fig. \ref{fig: Phi}.}
\label{fig: pat}
\end{figure}

Using similar arguments, we may already expect that for $\nu=7/5$ and $\nu=7/3$ lattice instabilities occur, since 
now the (mean-field) Landau levels are filled fractionally. 
Indeed we find them numerically, but they are of two different types. For $\nu=7/5$, we obtain two different self-consistent patterns (depending on the initial condition of the RMFT) with small but non-negligible energy difference and 
we concentrate on the one with lower energy first. As shown in Fig.~\ref{fig: pat}, remarkably,
the $\nu=7/5$ state does not exhibit charge modulation and has a uniform current amplitude on all bonds. However, the current pattern displays a $2\times 2$ plaquette modulation, with two plaquettes carrying opposite current loops and two plaquettes with zero current circulation. This is also correlated with
a $2\times 2$ modulation of the hopping $\chi_{ij\sigma}$. In contrast, the RMFT solution with higher energy (corresponding to a local minimum in the variational space) bears a more complicated bond structure. For $\nu=7/3$, CDW order along with BDW order always develops as shown in Fig.~\ref{fig: pat}. Interestingly, both cases can also be solved by including a non-zero pairing order parameter, indicating that either the Fermi level crosses bands instead of lying in a gap, or the gap is rather small compared to the cases of $\nu=1$ or $\nu=7$. Hence, superconductivity appears, as has been discussed before \cite{Poilblanc2}, coexisting with bond and/or charge orders.

Note that to find translation symmetry breaking states in the model, ED cannot be used since in our case its applicability is limited to a $4 \times 4$ cluster. For such a small system, finite size effects destroy the translational invariance even of non-interacting magnetic models. This is due to the gauge choice we have to make in order to implement a magnetic flux $\Phi = q/16, q = 0,\cdots,15$, which necessarily breaks the translational invariance within a $4 \times 4$ cluster. Of course, gauge invariance requires the full model to be translationally invariant. In the single particle picture, this can be accomplished by including degenerate states at nonzero momenta into the consideration. However, for the many-body system we are interested in, the system size accessible to ED is too small for these finite momentum single-particle states to contribute to the available Fock space. It is also not possible to effectively increase the system size by twisted boundary conditions as in the non-interacting case since this only ever allows us to reach a subset of all possible many-particle momenta: there are always many-particle momenta which correspond to different particles lying in different sectors of inserted flux, but twisted boundary conditions imply the same twisted flux for all particles. These shortcomings of ED render the comparison of charge, hopping and current density expectation values with RMFT difficult.

\section{\label{sec:FP}Fully polarized electron systems}\label{sec:fully polarized}

In the previous section we have considered a fixed doping of the $\rho=1/2$ Mott (AF) phase
and studied how states evolve with changes in flux. 
In this section we will now vary the electron density while setting $\Phi$ to be $7/16$ or $5/16$. The remarkable phenomenon 
discussed here is the instability towards a fully polarized ferromagnet where all electronic spins are aligned in the same direction. 
This instability is driven by a gain of kinetic energy happening in the ferromagnetic state which supersedes the loss of 
magnetic energy when the electron density
is small enough.  We have indeed found that the energies of fully polarized states are lower than those of the singlets, both in RMFT and ED, for a number of cases, and we shall focus on those in this section.

\begin{figure}[t]
\centering
\includegraphics[width=0.5 \textwidth]{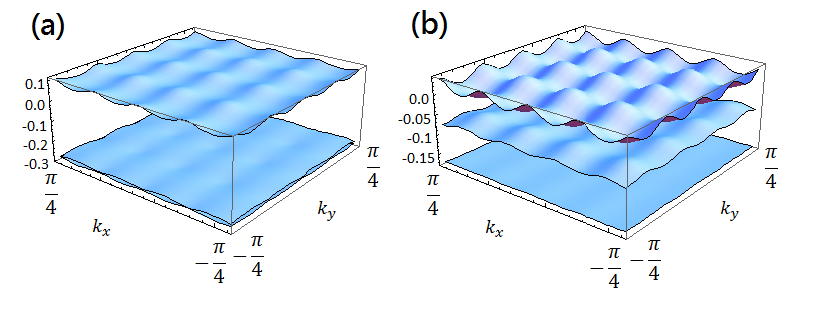}
\caption{Band structure for the three lowest energy bands for (a) $\nu^*=2/7$ and (b) $\nu^*=2/5$. At this doping, the first two bands are filled. Note that in (a) the first two bands are almost degenerate.}
\label{fig: bandstruct}
\end{figure}

For $\Phi=7/16$, we have studied several doping levels. For the cases we have considered, we found that the energies as calculated by RMFT or ED are very close (see Appendix) and the states we have found by either method are quite similar. This is not surprising since in fully polarized systems double occupancy is excluded by fermionic statistics, so that the projection operator $P_{\mathrm{G}}$ is no longer needed. Therefore, the Hamiltonian maps to a spinless electron system with nearest-neighbor repulsion. In this case, the RMFT renormalization factors given in the Appendix become 1 as expected. Note that this is obtained only if the variational parameters of the nearest neighbor sites are included in the expression of the renormalization factors~\cite{Himeda,Ogata,Yang3,Christensen} (small deviations from $1$ occur nevertheless for $g_{ij}^{s,z}$). The agreement between RMFT and ED asserts the reliability of RMFT in the low-electron density regime, far away from the widely investigated low-doping regime. To further confirm this, we have also made the comparison for the case of $\rho=1/16$ and $\Phi=5/16$ and the energies from both side still agree remarkably well. All states we have obtained possess only very small currents, meaning that the phases of $\chi_{ij}$ tend to screen the phases from the applied magnetic flux in order to lower the kinetic energy. However, for $\rho=1/16$ and $\Phi=5/16$, there also emerge CDW and BDW orders which are not seen for $\Phi=7/16$. This follows from the differences in the respective non-interacting band structures. In Fig.~\ref{fig: bandstruct}(a) for $\rho=1/16$ and $\Phi=7/16$ ($\nu^*=2\nu=2/7$), the Fermi level is located inside a large band gap between the second and the third (mean-field) band, producing a completely insulating state. 
In contrast, in Fig.~\ref{fig: bandstruct}(b) for $\rho=1/16$ and $\Phi=5/16$ ($\nu^*=2\nu=2/5$), the band gap is much smaller 
(for the $k$ points where the two consecutive bands are closest, the gap value is around $0.03 t$), which allows for the instabilities that have been observed in our calculation. 

\section{\label{sec:level2}Topological properties}\label{sec:FQH}

Together with charge/bond ordering, it is also particularly interesting to look for the emergence of FQH-type states with topological order. 
At half-filling ($\rho=1/2$) topological chiral spin liquids have been constructed as Gutzwiller projections of (non-interacting) wavefunctions
with a completely filled band of Chern number $\pm 1$~\cite{Wen1989,Laeuchli2015,Laeuchli2017}.
A related construction of topologically ordered states may also apply away from half-filling, at low doping and/or low electron density, 
and may be captured by the RMFT treatment of the Gutzwiller projector. In that case, our approach could point to situations where it may be energetically favorable for the system to accommodate a topologically ordered ground state.

Our first conclusion is that the $\nu=1$ and $\nu=7$ states in the integer quantum Hall regime are so robustly gapped that it is unlikely that further instabilities towards topologically ordered phases appear. 
What is left are the fully polarized uniform states with $\Phi=7/16$. 
The simplest prerequisite for the numerical realization of a FQH state in a system with periodic boundary conditions (i.e., a two-torus) is a topological ground state degeneracy (GSD).~\cite{Wen} In a given symmetry sector we expect nearly degenerate states which are separated by a gap from all other states. (If a system realizes a bosonic $\nu=1/2$ Laughlin state, this topological degeneracy should be two, for example.) 
Figure~\ref{fig: spectra} in the Appendix shows the ED energy spectra for each case that we have discussed, resolved into $S_z$ subspaces. We can see that there is no GSD even though for certain $S_z$ the first two energy levels are fairly close. For example, for $\nu=2/4$ the $S_z=0$ sector has two nearly degenerate states at low energy, but one has $S=0$ and the other one has $S=2$. Therefore, these states cannot be topologically degenerate partners. Moreover, we checked that the manifold spanned by these two states has even Chern number and thus cannot realize a FQH state.

The reason why it is hard for fully polarized phases to realize a FQH state in the model we study is as follows: The dominant Hubbard interaction term is very local. In the FQH effect, interaction terms, projected into the single particle states of a given Landau level, are expanded in Haldane pseudopotentials. An ultralocal interaction contributes to the $V_0$ pseudopotential, which gives rise to the bosonic Laughlin state. For the simplest fermionic FQH Laughlin state, the longer-ranged  pseudopotential  $V_1$ is required. However, as has been studied in the context of fractional Chern insulators \cite{Claassen}, the ultra-local Hubbard interaction translates into a dominant $V_0$ component after projection into a given band with nonvanishing Chern number.

\begin{table}[H]
\centering
\begin{tabular} {| c | c | c | c | c | c | c | c | }
\hline
$\rho$ & $\Phi$ & $\nu/\nu^*$ & $S$ & $C_{\rm ni} $ & $C_{\rm RMFT}$ & $C_{\rm ED}$ \\ \hline
\hline
7/16 & 		$7/16$ & 	1 & $0$ & 2 & 2 & 2 \\ \hline
7/16 & 	$5/16$ & 	$7/5	$ & $0$ &10 & 2  & 6\\ \hline
7/16 & 	$3/16$ & 	$7/3	$ & $0$ & 6 & 4  & 6 \\ \hline
7/16 & 		$1/16$ & 	$7	$ & $0$ & 14 & 2 & 14\\ \hline
7/32 & 	$7/16$ & 	$1^*$ & $FP$ & 1 & 1 &1 \\ \hline
1/8 & 	$1/4$ & 	$1^*$ & $FP$ & 1& 1  &1 \\ \hline
1/8 & 	$7/16$ & 	$4/7^*$ & $FP$ &  4& 4 & 4 \\ \hline
$1/16$ & 	$5/16	$ & $2/5^*$ & 	$FP$ & 6 &6  & 6 \\ \hline
$1/16$ & 	$7/16	$ & $2/7^*$ & $FP$ & 2 &  2 & 2 \\ \hline
\end{tabular}
\caption{\label{tab:Chern} Table comparing the Chern numbers obtained in the non-interacting case, in the (non-superconducting)
RMFT self-consistent solutions
 and by Lanczos ED. In the two first cases, the Chern numbers are given by summing up the contribution from all the filled bands. 
The last five rows noted by an asterisk represent the fully polarized states for which
$\nu^*=2\nu$ is listed instead of $\nu$.}
\end{table}

Although directly observing FQH states in our calculations seems therefore unlikely, the states we have obtained still have (generically) interesting topological features associated to non-zero {\it integer} Chern numbers~\cite{Thouless} and  Hall conductance given by
\begin{equation}
\begin{aligned}
\sigma=C \frac{e^2}{h}
\end{aligned}
\end{equation}
with $C$ being the (many-body) Chern number, and the Planck constant $h$ and the electronic charge $e$ 
have been re-introduced for clarity. For RMFT, the way of calculating Chern numbers is to integrate the Berry curvature of each mean-field band as has been shown in Ref.~\cite{Green}. In ED the many-body Chern numbers~\cite{Niu} are computed by introducing  twisted boundary conditions~\cite{Poilblanc1991,Fukui} 
(see Appendix for details).
The Chern numbers obtained by ED and RMFT (for the non-superconducting solutions) are compared with each other and also with the non-interacting case in Table~\ref{tab:Chern}. We note that at low enough electron filling, i.e., below $1/4$-filling, 
all Chern numbers agree with the non-interacting ones (provided one assumes a ferromagnetic state, e.g., 
considers spinless fermions) showing that the effect of the interaction is moderate in this 
regime. In particular we observe that the lattice 
instabilities found in RMFT do not affect the topological character of the states. 
In contrast, discrepancies appear when approaching the Mott insulating phase, in the low doping regime at $\rho=7/16$.
This signals that interactions play a crucial role there and obtaining the correct many-body Chern numbers of these correlated 
states is tedious: on one hand, the approximate way of treating the Gutzwiller projection in RMFT may not capture correctly the 
topological properties and/or, on the other hand, finite size effects in ED may also lead to deviations. It is, however, likely that
Chern numbers close to the Mott insulating phase are different from those of the non-interacting case. A noticeable counterexample is the case $\rho=\Phi=7/6$, $\nu=1$ where the Chern number $C=2$ obtained by ED and RMFT agrees with the non-interacting limit.
This suggests an adiabatic continuity from the non-interacting to the interacting case, which we have explicitly checked to hold in ED
using a Hofstadter-Hubbard model where we increased the interaction strength $U$ gradually.

Interestingly, it is possible to induce a transition from 
a topologically non-trivial phase to a trivial phase by adding a staggered potential
to the Hamiltonian, as was implemented in a cold atom 
experiment~\cite{Aidelsburger2}.
The staggered potential of magnitude $\Gamma$ takes the form:
\begin{equation}
H_{\rm staggered}=\frac{\Gamma}{2}\sum_{i}[(-1)^{i_x}+(-1)^{i_y}]n_{i}
\end{equation} 
where $n_{i}= c^\dagger_i c_i$ at lattice sites  $i=(i_x,i_y)$.  Notice that since we are considering fully polarized systems, we discard the spin index. Since the staggered potential has a $2\times 2$ spatial periodicity, it will induce CDW modulation via linear response, which may prohibit the formation of a topological phase (associated to a non-zero Chern number). To match the experimental setup, 
we choose here $\rho=1/8$ 
and $\Phi=1/4$, which gives $\nu^*=2\nu=1$. This corresponds to the scenario of a completely filled lowest Hofstadter band. The magnetic gauge used is shown in Fig. \ref{fig: Phi}(e). 
Our aim is to investigate the role of the interaction namely, (i) whether it could induce a lattice instability 
involving {\it spontaneous} translation  symmetry breaking and/or (ii) whether it will affect the location of the transition.
 
To investigate (i) we have used a $4\times 4$ supercell, larger than the $2\times 2$ magnetic unit cell, when solving the RMFT equations. In fact, no such instability was found, i.e., the $2\times 2 $ unit cell corresponds to the translation symmetry of the ground state.

We have considered different staggered potential strengths and observed the phase transition as a function of $\Gamma$
showing results very similar to the non interacting case~\cite{Aidelsburger,Aidelsburger2}
(The mean field band structure is shown in the Appendix). A qualitative understanding of the effect of the interaction on the location of the transition can be
obtained as follows. The mean-field Hamiltonian is in fact identical to the non-interacting one up
to renormalizations of the hopping term ${\tilde t} \approx t(1+\frac{J}{4}\chi)$  and of the magnitude of the $2\times 2$
potential ${\tilde\Gamma}=\Gamma(1+J\chi_{\Gamma})$, where the $J\chi_\Gamma\Gamma$ term originates from the induced 
effective local chemical potential (Eq. \ref{eq:che}) whose spatial periodicity is (in linear response) identical to the one of the perturbation $\Gamma$, and $\chi_{\Gamma}$ is a susceptibility at momentum $(\pi/2,\pi/2)$. 
As shown in Ref.\cite{Aidelsburger2}, the transition for a non-interacting system occurs at $\Gamma=2t$, which
for the mean-field Hamiltonian translates into ${\tilde\Gamma}=2{\tilde t}$ providing a simple expression for the critical staggered
strength $\Gamma^*$,
\begin{equation}
\Gamma^*=2t\,\,\frac{1+\frac{1}{4}J\chi}{1+J\chi_{\Gamma }}.\end{equation} 
Using the numerical values of $\chi$ (Eq.~\ref{eq:chi}) and $\chi_\Gamma$ at $J=0.3$, we obtain $\Gamma^*\simeq 2.048t$.
This signifies that interactions increase the size of the trivial region only very slightly, which may be a generic feature.

\section{\label{sec:level2}conclusion}
\label{sec:conclusion}

Motivated by recent experimental and numerical developments, we studied the Harper-Hofstadter model in the presence of strong correlations, which corresponds to the $t$--$J$ model in an orbital magnetic field. By employing a RMFT approach, 
supplemented by Lanczos ED calculations, we endeavored to find novel condensed matter phases for fermionic systems.  
In particular, we have focused on CFPs and several ferromagnetic phases. 
Although we failed to observe topologically ordered states, neither of singlet character nor fully polarized,
topologically non-trivial states with non-zero Chern numbers have been identified in the presence of interaction. 
We found CFPs which realize an integer quantum Hall system. Those at fractional filling fraction $\nu$ generically exhibit lattice instabilities. For fully polarized states, occurring at low electron filling, RMFT and ED agree precisely with each other with regard to the GS energies and Chern numbers. Moreover, we showed that
the effect of a staggered potential on destabilizing the topological state depends weakly on the interaction. 
Note that, close to the Mott insulating phase, i.e., at low (hole) doping, RMFT and ED results for the Chern numbers disagree with each other, revealing strong interaction effects that render the computation of the topological properties of the states difficult. Therefore, it is interesting to realize the system we propose in experimental setups. It has been shown that it is possible to investigate the Fermi-Hubbard model with degenerate Fermi gases with atomic species such as $^{6}$Li(37, 38), manipulated within optical lattices \cite{Esslinger}. In order to include (synthetic) gauge fields, laser assistant tunneling can be applied with two laser beams controlling the hopping of nearby sites with an additional flux phase \cite{Aidelsburger, Miyake}. We suggest a combination of these techniques for an experimental investigation of our system.
Compared with the experimental setup of the Harper-Hofstadter Hamiltonian with interaction, however, the agreement between our results and those from the cold atom experiment suggests that the $t$--$J$ Hamiltonian is relevant for describing the physics of interacting fermions under external magnetic flux.
Our results give a taste of the phenomena emerging from the strongly correlated Hofstadter Hamiltonian and motivate further experimental and theoretical studies. 

\begin{acknowledgements}
This project is supported by the TNSTRONG
ANR grant (French Research Council).  This work was granted access to the HPC resources of CALMIP 
supercomputing center under the allocation 2017-P1231. 
F.S. and T.N. acknowledge support from the Swiss National Science foundation under grant 200021\_169061.
\end{acknowledgements}


\newpage
\bibliography{paper_draft2}

\newpage
\appendix

\section{\label{sec:level1}Renormalized Mean Field Theory}

\begin{figure}[t]
\centering
\includegraphics[width=0.5 \textwidth]{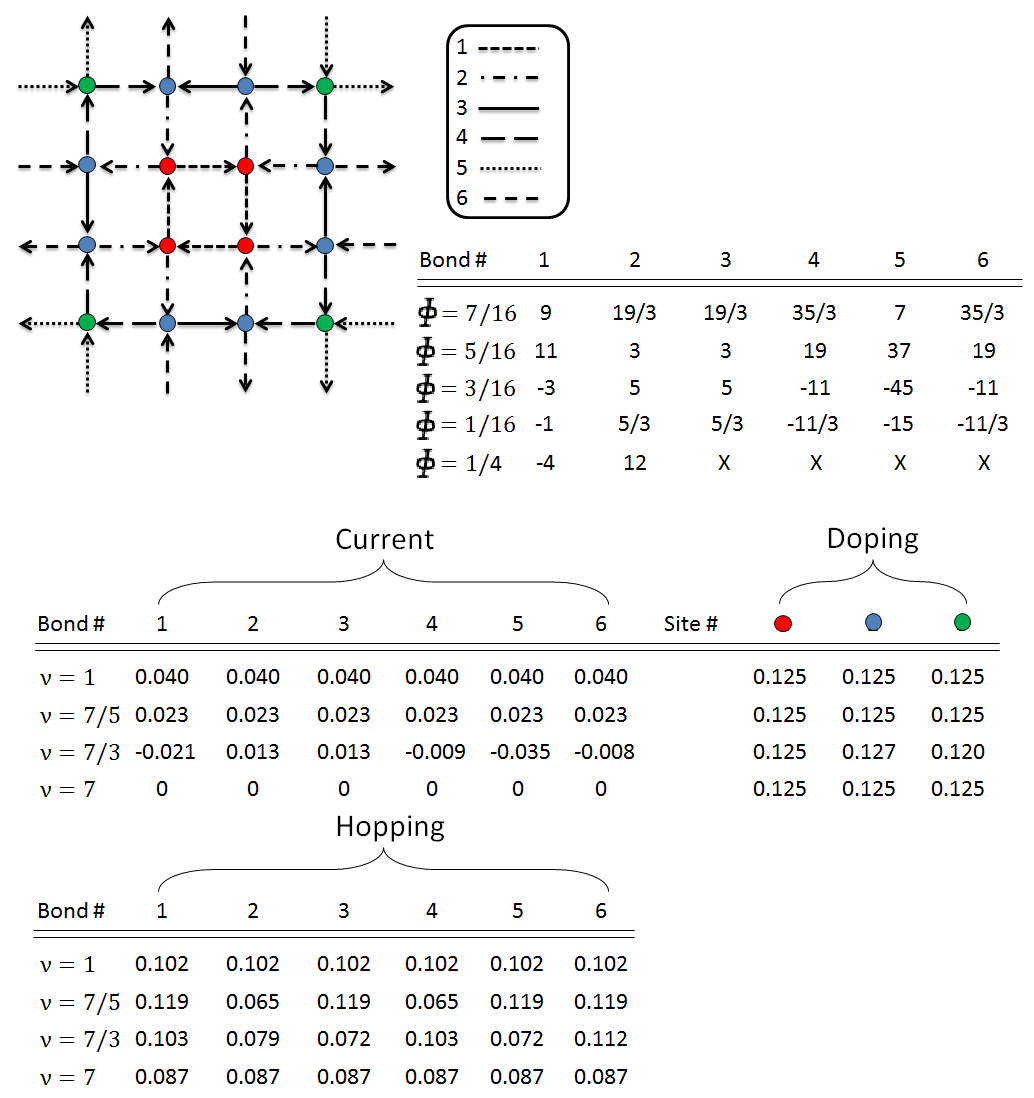}
\caption{Distribution of the phases $\phi_{ij}$ on the bonds of $4\times 4$ and $2\times 2$ unit cells (on the 2-torus) for the
flux densities $\Phi$ considered in this work. Arrows again indicate the directions of current and negative signs stand for opposite flows. The flux density $\Phi=1/4$ has only two different bonds (bond 1 and 2). The lower panel shows detailed numbers of variables for the patterns in Fig. \ref{fig: pat}.}
\label{fig: Phi}
\end{figure}

In order to deal with the projection operators $P_{\mathrm{G}}$ in Eq.~\ref{eq:H} for RMFT, we replace them by Gutzwiller renormalization factors. The renormalized Hamiltonian now reads
\begin{equation}
\begin{aligned}
\label{eq:H2}
H=&-\sum_{\langle i,j\rangle\mu}g^t_{ij\mu}t_{ij}e^{i\phi_{ij}}(c^\dagger_{i\mu}c_{j\mu}+\mathrm{h.c.})\\
&+\sum_{\langle i,j\rangle}J\Bigg [ g^{s,z}_{ij}S^{s,z}_i S^{s,z}_j+g^{s,xy}_{ij}\Bigg(\frac{S^{+}_i S^{-}_j+S^{-}_i S^{+}_j}{2}\Bigg)\Bigg] \\
\end{aligned}
\end{equation}
where $g^t_{ij\sigma}, g^{s,z}_{ij}$, and $g^{s,xy}_{ij}$ are the Gutzwiller factors, which depend on the values of the pairing field $\Delta_{ij\mu}^v$, bond order $\chi_{ij\mu}^v$, and hole density $\delta_i$: 
\begin{equation}
\begin{aligned}
\label{eq:P}
&m_i^v=\langle\Psi_0 | S^z_i | \Psi_0 \rangle\\
&\Delta_{ij\mu}^v=\mu \langle\Psi_0 | c_{i\mu}c_{j\bar{\mu}} | \Psi_0 \rangle\\
&\chi_{ij\mu}^v=\langle\Psi_0 | c^\dagger_{i\mu}c_{j\mu} |\Psi_0 \rangle\\
&\delta_i=1-\langle\Psi_0 | n_i |\Psi_0 \rangle
\end{aligned}
\end{equation}
where $| \Psi_0 \rangle$ is the unprojected wavefunction. The superscript $v$ is used to denote that these quantities are variational parameters instead of real physical quantities. As for the phases ($\phi_{ij}$), we followed Ref.~\cite{Poilblanc}. The numbers for different flux per plaquette $\Phi$ are shown in Fig. \ref{fig: Phi}. We will start by considering the Gutzwiller factors first proposed by Ogata and Himeda \cite{Himeda,Ogata}, which are given by
\begin{equation}
\begin{aligned}
\label{eq:G}
&g^t_{ij\mu}=g^t_{i\mu}g^t_{j\mu}\\
&g^t_{i\mu}=\sqrt{\frac{2\delta_i(1-\delta_i)}{1-\delta_i^2+4(m_i^v)^2}\frac{1+\delta_i+\mu 2m_i^v}{1+\delta_i-\mu 2m_i^v}} \\
&g^{s,xy}_{ij}=g^{s,xy}_i g^{s,xy}_j\\
&g^{s,xy}_i=\frac{2(1-\delta_i)}{1-\delta_i^2+4(m_i^v)^2}\\
&g^{s,z}_{ij}=g^{s,xy}_{ij} \frac{2((\bar{\Delta}^v_{ij})^2+(\bar{\chi}^v_{ij})^2)-4m_i^vm_j^vX^2_{ij}}{2((\bar{\Delta}^v_{ij})^2+(\bar{\chi}^v_{ij})^2)-4m_i^vm_j^v}\\
&X_{ij}=1+\frac{12(1-\delta_i)(1-\delta_j)((\bar{\Delta}^v_{ij})^2+(\bar{\chi}^v_{ij})^2)}{\sqrt{(1-\delta_i^2+4(m_i^v)^2)(1-\delta_j^2+4(m_j^v)^2)}}
\end{aligned}
\end{equation}
where $\bar{\Delta}^v_{ij}=\sum_{\mu}\Delta_{ij\mu}^v/2$ and $\bar{\chi}^v_{ij}=\sum_{\mu}\chi_{ij\mu}^v/2$. For singlet states the magnetization $m^v_i$ is equal to zero and $n_{i\uparrow}=n_{i\downarrow}=\frac{1}{2}(1-\delta_{i})$. However, for the fully polarized scenario $m^v_i=n_{i\uparrow}/2$ while $n_{i\uparrow}=(1-\delta_i)$, $n_{i\downarrow}=0$, where we assume that all electrons have spin up. This set of Gutzwiller factors corresponds to finite doping and is consistent with variational Monte Carlo calculations \cite{Himeda, Ogata}. 

After we replace the projection operator by the Gutzwiller factors by using the mean-field order parameters defined in Eq.~\ref{eq:P} and ~\ref{eq:G}, the energy of the renormalized Hamiltonian(Eq.~\ref{eq:H2}) becomes 
\begin{equation}\label{first}
\begin{aligned}
E=\langle\Psi_0 \mid H \mid\Psi_0 \rangle=&-\sum_{i,j,\mu}g^t_{ij\mu}te^{i\phi_{ij}}(\chi_{ij\mu}^v+\mathrm{h.c.})\\
&-\sum_{\langle i,j \rangle\mu}J\Big (\frac{g^{s,z}_{ij}}{4}+\frac{g^{s,xy}_{ij}}{2}\frac{\Delta^{v\ast}_{ij\bar{\mu}}}{\Delta^{v\ast}_{ij\mu}}\Big )\Delta^{v\ast}_{ij\mu}\Delta^v_{ij\mu}\\
&-\sum_{\langle i,j \rangle\mu}J\Big (\frac{g^{s,z}_{ij}}{4}+\frac{g^{s,xy}_{ij}}{2}\frac{\chi^{v\ast}_{ij\bar{\mu}}}{\chi^{v\ast}_{ij\mu}}\Big )\chi^{v\ast}_{ij\mu}\chi^v_{ij\mu}\\
&+\sum_{\langle i,j \rangle}g^{s,z}_{ij}Jm^v_im^v_j  
\end{aligned}
\end{equation} 
where $m^v_i$ represents the spin moments, here we set them to be zero for all of the following cases.

Next we want to minimize the energy under two constraints: $\sum_in_i=N_{\mathrm{e}}$ and $\langle\Psi_0 | \Psi_0 \rangle=1$. Thus our cost function to be minimized is
\begin{equation}
W=\langle\Psi_0 | H |\Psi_0 \rangle-\lambda(\langle\Psi_0 | \Psi_0 \rangle-1)-\epsilon\big(\sum_in_i-N_{\mathrm{e}}\big) 
\end{equation} 
The mean-field Hamiltonian becomes
\begin{equation}
\begin{aligned}
\label{eq:MF}
H_{\mathrm{MF}}&=\sum_{\langle i,j\rangle\mu}\frac{\partial W}{\partial \chi^v_{ij\mu}}c^\dagger_{i\mu}c_{j\mu}+\mathrm{h.c.}\\
&+\sum_{\langle i,j \rangle\mu}\frac{\partial W}{\partial \Delta^v_{ij\mu}}\mu c_{i\mu}c_{j\bar{\mu}}+\mathrm{h.c.}\\
&+\sum_{i,\mu}\frac{\partial W}{\partial n_{i\mu}}n_{i\mu}
\end{aligned}
\end{equation} 
Eq.~(\ref{eq:MF}) satisfies the Schr\"odinger equation $H_{\mathrm{MF}}| \Psi_0 \rangle=\lambda| \Psi_0 \rangle$. The three derivatives are defined as 
\begin{equation}
\begin{aligned}
H_{ij\mu}=\frac{\partial W}{\partial \chi^v_{ij\mu}}=&-J\Big (\frac{g^{s,z}_{ij}}{4}+\frac{g^{s,xy}_{ij}}{2}\frac{\chi^{v\ast}_{ij\bar{\mu}}}{\chi^{v\ast}_{ij\mu}}\Big )\chi^{v\ast}_{ij\mu}\\
&-g^t_{ij\mu}t_{ij}e^{i\phi_{ij}}+\frac{\partial W}{\partial g^{s,z}_{ij}}\frac{\partial g^{s,z}_{ij}}{\partial \chi^v_{ij\mu}}\\
\end{aligned}
\label{eq:chi}
\end{equation} 

\begin{equation}
\begin{aligned}
D^\ast_{ij}=\frac{\partial W}{\partial \Delta^v_{ij\uparrow}}=&-J\Big (\frac{g^{s,z}_{ij}}{4}+\frac{g^{s,xy}_{ij}}{2}\frac{\Delta^{v\ast}_{ij\downarrow}}{\Delta^{v\ast}_{ij\uparrow}}\Big )\Delta^{v\ast}_{ij\uparrow}\\
&+\frac{\partial W}{\partial g^{s,z}_{ij}}\frac{\partial g^{s,z}_{ij}}{\partial \Delta^v_{ij\uparrow}}\\
\end{aligned}
\end{equation} 

\begin{equation}
\begin{aligned}
\epsilon_{i\mu}=-\frac{\partial W}{\partial n_{i\mu}}=&\epsilon-\sum_j\frac{\partial W}{\partial g^{s,xy}_{ij}}\frac{\partial g^{s,xy}_{ij}}{\partial n_{i\mu}}\\
&-\sum_j\frac{\partial W}{\partial g^{s,z}_{ij}}\frac{\partial g^{s,z}_{ij}}{\partial n_{i\mu}}-\sum_{j\mu'}\frac{\partial W}{\partial g^t_{ij\mu'}}\frac{\partial g^t_{ij\mu'}}{\partial n_{i\mu}}
\end{aligned}
\label{eq:che}
\end{equation} 
Eq.~(\ref{eq:che}) is the effective local chemical potential.

Now $H_{\mathrm{MF}}$ can be rewritten in form of the BdG equations
\begin{equation}
\begin{aligned}
H_{\mathrm{MF}}= 
\left(
	c^\dagger_{i\uparrow},c_{i\downarrow}
\right)
\left(
	\begin{array}{cc}
	H_{ij\uparrow} & D_{ij}\\
	D^\ast_{ji} & -H_{ji\downarrow}\\
	\end{array}
\right)
\left(
	\begin{array}{cc}
	c_{j\uparrow} \\
	c^\dagger_{j\downarrow} \\
	\end{array}
\right)
\end{aligned}
\end{equation}
We can diagonalize $H_{\mathrm{MF}}$ to obtain an equal number of positive and negative eigenvalues together with their corresponding eigenvectors $(u^n_i,v^n_i)$. Then we can make use of the eigenfunctions we have got for the following iteration until self-consistency is achieved.

For each band its Chern number is defined by integrating the Berry curvature over the first Brillouin zone \cite{Green}:

\begin{equation}
\begin{aligned}
&C_n=\frac{1}{2\pi}\sum_{\textbf{k}\in BZ}\nabla_{\textbf{k}}\times \vec A_n(\textbf{k})=\frac{1}{2\pi}\sum_{\textbf{k}\in BZ}\vec B_n(\textbf{k})\\
&=\frac{-i}{2\pi}\sum_{m \neq n}\,\sum_{\textbf{k}\in BZ}\frac{\big< n\textbf{k}|J_x|m\textbf{k}\big>\big< m\textbf{k}|J_y|n\textbf{k}\big> - (J_x \leftrightarrow J_y)}{[E_n(\textbf{k})-E_m(\textbf{k})]^2} \, \\
\end{aligned}
\end{equation}
where $A_n(\textbf{k})=-i\big< n\textbf{k}|\nabla_{\textbf{k}}|n\textbf{k} \big>$ is the Berry vector field for the nth band, and $\vec B_n(\textbf{k})$ is the related field. The current $\textbf{J}=(J_x,J_y)$ is given by $J=\nabla_{\textbf{k}}H$. 

\begin{table}[H]
\centering
\begin{tabular} {| c | c |  c | c | c | c | c | c | c |}
\hline
   $\rho$  & $\Phi$ & $\nu/\nu^*$& $E_0$ &$E_{\rm kin}$ &$E_{\rm pot}$ &$C_{\rm RMFT}$ \\ 
\hline
7/16 & $7/16$& $1$ &-8.945t    & -6.539t & -2.405t &2\\
\hline
7/16  &$5/16$ &$7/5$&-8.119t  & -5.882t & -2.238t &2\\
\hline
 7/16& $3/16$&$7/3$ &-7.632t & -5.616t & -2.016t &4\\
\hline
7/16& $1/16$&$7$ & -7.658t  & -5.562t & -2.096t &2\\
\hline
7/32&$7/16$ &$1^*$  &-14.353t  & -14.713t & 0.360t & 1\\
\hline
1/8& $1/4$&$1^*$ & -10.834t &-10.917t & 0.083t & 1 \\
\hline
1/8&$7/16$ &$4/7^*$  &-9.467t  & -9.566t & 0.098t &4\\
\hline
1/16& $5/16$&$2/5^*$  & -5.253t  & -5.274t & 0.021t &6\\
\hline
 1/16&$7/16$ &$2/7^*$ & -5.176t  & -5.197t & 0.022t &2\\
\hline
\end{tabular}
\caption{\label{tab:example}Table of the energies and Chern numbers for the self-consistent solutions obtained in RMFT. $E_0=E_{\rm kin}+E_{\rm pot}$ represents the energy per $4\times 4$ sublattice. The last column is the Chern number given by summing up the contribution from all the filled (mean-field) bands. 
The last five rows noted by an asterisk represent the fully polarized states for which
$\nu^*=2\nu$ is listed instead of $\nu$.
}
\end{table}

\section{\label{sec:level1}Exact Diagonalization}

\subsection{\label{sec:level1}Model}
\begin{figure}[t]
\centering
\includegraphics[width=0.5 \textwidth]{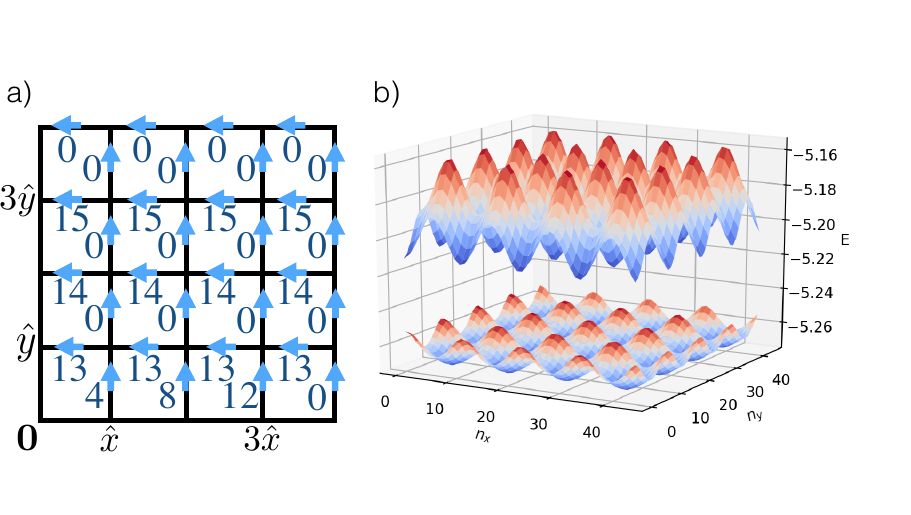}
\caption{(a) Vector potential gauge choice for $\Phi = q/16, q = 0,\cdots,15$. Periodic boundary conditions are assumed. $A_{ij}$ in units of 
$F=2\pi\Phi$ is given by the integer number shown between site $i$ and $j$, with positive sign if the respective arrow points from site $i$ to site $j$, and negative sign otherwise.
(b) Spectrum $E(\bs{\phi})$ as a function of inserted flux for $\nu=1/5$. The Chern number evaluates to $6$, however, there is no indication for a topological GSD.
}
\label{fig: gauge}
\end{figure}

We study by Lanczos ED an instance of the model given by Eq.~\eqref{eq:H} with $\Phi = q/16, q = 0,\cdots,15$, for the parameter $t=1$ and $J=0.3$, on a $4 \times 4$ lattice with periodic boundary conditions (2-torus geometry). We make a choice of gauge in which the $A_{ij}$ take the values shown in Fig.~\ref{fig: gauge}(a).

$H$ preserves the total number of particles per spin $n_\mu = \sum_{i,\mu} n_i$, which is therefore a good quantum number. 
For one, this allows us to treat sectors of different particle number. 
We will label them by the particle filling $\rho = (n_\uparrow + n_\downarrow)/32$.
On the other hand, the model is also invariant under global $SU(2)$ spin rotations. In particular, it is unaffected by global $U(1)$ rotations around the $z$-axis. The eigenvalue of the operator $S_z = \sum_i (S_z)_i$ is therefore a good quantum number, and we can diagonalize $H$ in each $S_z$ subspace separately. Finite-size precursors to ferromagnetic order can be inferred from degenerate energy eigenvalues at different $S_z$, where a multiplicity of $2S+1$ corresponds to a spin polarization of magnitude $S$.

\subsection{Many-body Chern number}
To calculate the many-body Chern number, we introduce twisted boundary conditions~\cite{Poilblanc1991} labeled by the twisting angles $\bs{\phi} = (\phi_x,\phi_y)^\mathsf{T}$. This amounts to all many-body states $\ket{\Psi}$ obeying
\begin{equation}
T_{L \hat{x}} \ket{\Psi} = e^{\mathrm{i} \phi_x} \ket{\Psi}, \quad
T_{L \hat{y}} \ket{\Psi} = e^{\mathrm{i} \phi_y} \ket{\Psi},
\end{equation}
where $T_{\bs{r}}$ is any operator translating a single particle by $\bs{r}$. In practice, this prescription can be implemented by making the substitutions 
\begin{equation}
\begin{aligned}
A_{i,i+\hat{x}} &\rightarrow A_{i,i+\hat{x}} + \phi_x, \\ &\forall i = (L-1)\hat{x} + n \hat{y}, \quad n=0,\cdots,L-1, \\
A_{i,i+\hat{y}} &\rightarrow A_{i,i+\hat{y}} + \phi_y, \\ &\forall i = (L-1)\hat{y} + n \hat{x}, \quad n=0,\cdots,L-1.
\end{aligned}
\end{equation}

The Chern number of the $n$-th many body eigenstate $\ket{n}$ is then defined as~\cite{Niu}
\begin{equation}
C = \frac{1}{2 \pi \mathrm{i}} \int_0^{2\pi} d \phi_x \int_0^{2\pi} d \phi_y \epsilon^{a b} \braket{\partial_a n(\bs{\phi}) | \partial_b n(\bs{\phi})},
\end{equation}
where $\epsilon^{a b}$, $a, b = x,y$ is the totally antisymmetric $2 \times 2$ tensor, $\partial_a = \partial/\partial \phi_a$, and we assume that $\ket{n(\bs{\phi})}$ is non-degenerate at all $\bs{\phi}$.

In practice, to calculate the Chern number via ED, we consider a lattice of twisted boundary conditions $\phi_a = 2\pi \, n_a/N$, $n_a = 0...N-1$, and evaluate $C$ using the prescription of Ref.~\cite{Fukui}. Here, we have chosen $N=45$ for the cases corresponding to low fermion densities. For the cases corresponding to $\rho=7/16$ filling, 
i.e., 2 holes on $4\times 4$, which have a much larger Hilbert space,  
we have taken $N=10$ and checked the consistency of the results with $N=32$ in the special case where $\Phi=5/16$.
See Fig.~\ref{fig: gauge}(b) for an example of the dependence of the spectrum of $H$ on inserted flux.

\subsection{Results}
\begin{figure*}[t]
\centering
\includegraphics[width=1. \textwidth]{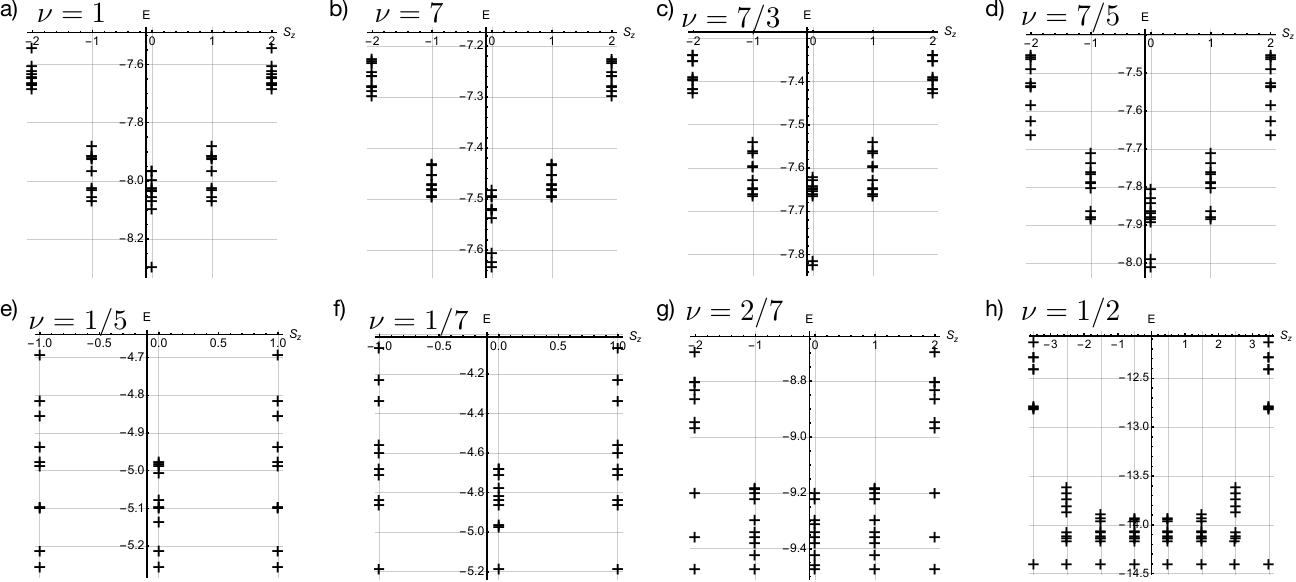}
\caption{Lanczos ED spectrum of $H$ for various values of $\nu$, with $\Phi$ and $\rho$ as given by Table~\ref{tab: MagTable}. When there is no magnetization, only the $S_z = 0, \pm 1$ sector is shown.
}
\label{fig: spectra}
\end{figure*}

We diagonalize $H$ for various filling factors $\nu$, defined as $\nu \equiv \rho / \Phi$. The GS energies, as well as spin polarizations and Chern numbers are summarized in Table~\ref{tab: MagTable}. Figure~\ref{fig: spectra} furthermore shows the spectra for the $S_z$ values of interest. Taking $\ket{0}$ to be the many-body GS of $H$, we define $E = \bra{0} H \ket{0}$, $E_{\text{kin}} = \bra{0} H_{\text{kin}} \ket{0}$ and $E_{\text{pot}} = \bra{0} H_{\text{pot}} \ket{0}$, with $H_{\text{kin}}$ and $H_{\text{pot}}$ given by Eq.~\eqref{eq:H}.

\begin{table}[H]
\centering
\begin{tabular} {| c | c | c | c | c | c | c | c | c |}
\hline
$\rho$ &$\Phi$ & $\nu/\nu^*$ &  $S$ & $E_0$ & $E_\text{kin}$ & $E_\text{pot}$ & $C_{\rm ED}$ \\ \hline
\hline
	$7/16$ & 	$7/16	$ &$1$ & 	 $0$ & $-8.2901$ & $-6.39644$ & $-1.89369$ &2  \\ \hline
$7/16	$ &$5/16$ & 	$7/5$ & 	 $0$ & $-8.0058$ & $-6.04586$ & $-1.95997$ &6 \\ \hline
$7/16	$ &$3/16$ & 	 $7/3$ & 	$0$ & $-7.8204$ & $-5.90818$ & $-1.91226$ &6 \\ \hline
$7/16	$ &	$1/16$ & 	$7$ & 	 $0$ & $-7.6298$ & $-5.73802$ & $-1.89179$ &14 \\ \hline
$7/32	$ &$7/16$ & 	$1^*$ & 	 $7/2$ & $-14.3874$ &$-14.7165$ & $0.329042$ &1 \\ \hline
$1/8	$ &$1/4$ & 	$1^*$ & 	 $2$ & $ -11.2393 $ &$-11.3132$ & $0.0739077$ &1 \\ \hline
$1/8	$ &$7/16$ & 	$4/7^*$ & 	 $2$ & $-9.4670$ & $-9.55201$ & $0.0849988 $ & 4 \\ \hline
$1/16	$ &$5/16$ & 	$2/5^*$ & 	 $1$ & $-5.2519$ & $-5.26527$ & $0.0133967 $ & 6 \\ \hline
$1/16	$ &	$7/16$ & 	 $2/7^*$ & $1$ & $-5.1794$ & $-5.19852$ & $0.0190752 $ & 2 \\ \hline
\end{tabular}
\caption{Summary of the Lanczos exact diagonalization results.}
\label{tab: MagTable}
\end{table}

\section{Induced topological-trivial transition}

\begin{figure*}[t]
\begin{center}
\includegraphics[width=1.0 \textwidth]{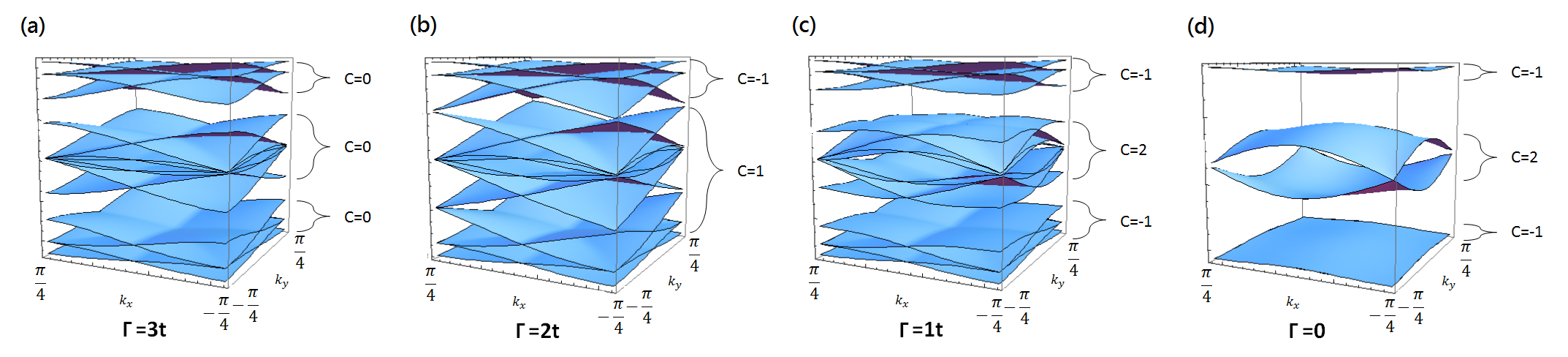}
\caption{RMFT energy spectrum as a function of staggered potential $\delta$ with Chern numbers for each band shown beside the figure. For $\Gamma>2t$ the system is topologically trivial with the Chern number $C$ of the bands zero. At the transition point, the band gap closes and it becomes topologically non-trivial with $C=1$ for the lowest band. After passing the transition point, the gap opens again and the lowest band now possesses a Chern number of -1. Notice that within this chosen reduced BZ, each of the four bands originating from the $2\times 2$ modulation is folded into 4 sub-bands, producing a total of 16 bands.}
\label{fig: 1/2}
\end{center}
\end{figure*}

Further details of the RMFT calculations of the Hofstadter $t$--$J$ model for $\rho=1/8$ and $\Phi=1/4$ and 
in the presence of a staggered potential are given here.
We have considered different staggered strengths and observed the phase transition described
in Ref.~\cite{Aidelsburger2} as a function of $\Gamma$. In Fig.~\ref{fig: 1/2}, the band structures for four representative values of $\Gamma$ are shown within a reduced Brillouin zone (BZ), $k_x,k_y \in [-\pi/4,\pi/4]$. Note that the modulations generated by a non-zero $\Gamma$ all have $2\times 2$ periodicity, indicating that the bands connecting with each other at the zone boundary are in fact due to the (artificial) band folding originating from the larger supercell used in the RMFT calculation (see main text), and therefore should be considered as the same bands. In Fig. \ref{fig: 1/2}(a), the bands are topologically trivial since their Chern numbers are zero. There is also an obvious band gap between the lowest and middle bands. As we lower the staggered value, the gap shrinks gradually and closes eventually at $\Gamma\simeq 2t$. This is the point when the system enters the topologically non-trivial phase since now summing up the Chern numbers of the lowest and middle bands gives 1. As we further lower the staggered strength, the gap opens up again and the Chern numbers for the highest, middle, and lowest bands become -1, 2, and -1, respectively. When the staggered number is equal to zero, the system is similar to the Harper-Hofstadter model with $\Phi=1/4$. Our results reveal a competition between the topological phase and the
(induced) CDW, which has been experimentally realized by Aidelsburger \textsl{et al.}\cite{Aidelsburger2}.

\end{document}